\documentclass[a4paper]{article}
\usepackage[numbers]{natbib}
\usepackage{amsmath, amssymb, amsfonts}
\usepackage{hyperref}
\usepackage{tabularx}
\usepackage{changepage}
\usepackage{algorithm}
\usepackage[noend]{algpseudocode}
\usepackage{enumitem}
\usepackage{multirow}
\usepackage{color}
\usepackage[labelfont=bf,font=small,margin=1mm]{caption}
\usepackage{hhline}
\usepackage{graphicx}
\usepackage[affil-it]{authblk}

\newcommand{\JITBP}{JIT with BP}
\newcommand{\OBJ}[1]{\ensuremath{{\texttt{WET}}\!\left(#1\right)}}
\newcommand{\pack}{\texttt{PACK}}
\newcommand{\AB}{ABH}
\newcommand{\CP}{CPH}
\newcommand{\GroupFormation}{\texttt{GroupFormation}}
\newcommand{\GroupJoin}{\texttt{GroupJoin}}
\newcommand{\lbfs}{lb_{f\!s}}
\newcommand{\eq}[2]{\ensuremath{{\texttt{Equals}}\!\left(#1,#2\right)}}
\newcommand{\el}[2]{\ensuremath{{\texttt{Element}}\!\left(#1,#2\right)}}
\newcommand{\fa}[2]{\ensuremath{#1 \!\in\! #2}}
\newcommand{\ad}[2]{\ensuremath{{\texttt{AllDifferent}}\!\left[#1,\ldots,#2\right]}}
\newcommand{\countn}[2]{\ensuremath{{\texttt{Count}}\!\left(#1,#2\right)}}
\newcommand{\formula}[2]{{\small\begin{flalign}#1\label{#2}\end{flalign}}}
\newcommand{\WETbin}[2]{\ensuremath{{\texttt{WET}}\!\left(#1,#2\right)}}
\newcommand{\overbar}[1]{\mkern 1.5mu\overline{\mkern-1.5mu#1\mkern-1.5mu}\mkern 1.5mu}

\begin{document}

\title{Just-in-Time Batch Scheduling Problem with Two-dimensional Bin Packing Constraints}

\author[1]{S. Polyakovskiy} 
\author[1]{A. Makarowsky} 
\author[2]{R. M'Hallah} 
{\footnotesize
\affil[1]{Optimisation and Logistics Group, School of Computer Science, University of Adelaide, Australia.}
\affil[2]{Department of Statistics and Operations Research, \newline College of Science, Kuwait University, Kuwait.}
}

\maketitle

\begin{abstract}
This paper introduces and approximately solves a multi-component problem where small rectangular items are produced from large rectangular bins via guillotine cuts. An item is characterized by its width, height, due date, and earliness and tardiness penalties per unit time. Each item induces a cost that is proportional to its earliness and tardiness. Items cut from the same bin form a batch, whose processing and completion times depend on its assigned items. The items of a batch have the completion time of their bin.  The objective is to find a cutting plan that minimizes the weighted sum of earliness and tardiness penalties. We address this problem via a constraint programming (CP) based heuristic ({\CP}) and an agent based modelling heuristic ({\AB}). {\CP} is an impact-based search strategy, implemented in the general-purpose solver IBM CP Optimizer. {\AB} is constructive. It builds a solution through repeated negotiations between the set of agents representing the items and the set representing the bins. The agents cooperate to minimize the weighted earliness-tardiness penalties. The computational investigation shows that {\CP} outperforms {\AB} on small-sized instances while the opposite prevails for larger instances.
\end{abstract}


\section{Introduction}

Multi-component problems are increasingly suscitating the interest of Evolutionary Computation (EC) and Operations Research (OR) communities \cite{Polyakovskiy2014TTP, Bonyadi2016}. Not only do they combine several combinatorial optimisation aspects into a single problem but they also emanate from the compounded complexity of conflicting issues in areas like logistics and supply chain management. Solving them requires a thorough understanding of both their compounded and their individual natures.  In practice, solving a multi-component problem is harder than separately tackling its components.

This paper focuses on a relevant multi-component problem occurring in make-to-order industries that adopt a pull production strategy. A production station requests some parts from preceding stages, and specifies when it needs each part. The preceding stages react by producing the parts on time in order to avoid handling, transfer, and temporary storage when parts are completed early and starving of subsequent stations and delays when parts are finished tardy. Synchronizing the production requires that every stage complete its workload on time.

The problem we consider involves the cutting of raw material at one of its production stages. It occurs in furniture, wood, and plastic industries.  The cutting stage produces a set $N = \left\{1,\ldots,n\right\}$ of $n$ small rectangular items from a set $B=\left\{1,\ldots,m\right\}$ of large identical rectangular sheets of raw material, referred to hereafter as bins. It uses a single guillotine cutting machine whose cuts are parallel to the edges of the sheets. That is, every item is obtained by a series of edge to edge parallel straight cuts. Item $i \in N$ is characterized by its width $w_i,$ height $h_i,$ due date $d_i,$ which defines when $i$ should be ideally produced, a per time unit earliness cost $\epsilon_i,$ and a per unit tardiness cost $\tau_i.$  Depending on the items' dimensions, it is possible to cut more than one item from a bin. A bin $k \!\in\! B$ is characterized by its width $\overbar{W}$ and height $\overbar{H}$. The number of bins $m \!\leq\! n.$ That is, at worst, each item is packed in a single bin. A subset $N_k \subseteq N$ of items assigned to a bin $k \in B$ can not overlap and should be completely contained in the bin. The subset forms a batch whose processing time is a function $f\left(N_k\right)$ of $N_k$ and whose completion time is $c_k.$  The completion time $c_i$ of item $i \in N_k$ equals $c_k.$ When produced earlier than $d_i$, item $i$ generates earliness $E_i = \max\left\{0,d_i - c_i\right\}$ yielding an earliness penalty $\epsilon_i E_i.$ On the other hand, it has a tardiness $T_i = \max\left\{0,c_i - d_i\right\}$ and a tardiness cost $\tau_i T_i$ when produced later than $d_i.$ This problem, denoted {\JITBP}, searches for (i) a feasible guillotine packing of $N$ into bins of $B$, and (ii) its bin cutting schedule that minimizes the total weighted earliness-tardiness $\textstyle\sum_{i \in N} \left(\epsilon_i E_i + \tau_i T_i\right)$.

Applications combining bin packing and scheduling range from steel manufacturing~\cite{Reinertsen10,Arbib14} to ship lock scheduling~\cite{Verstichel15} to wood cutting~\cite{Polyakovskiy11}.  However,
to the best of our knowledge, no research deals with {\JITBP} introduced in this paper. The problem is motivated by the specificity of make-to-order industries, which are characterised by low-volume high-mix orders with fixed due dates~\cite{plataine}. To be flexible, such industries don't produce large batches of identical items. Their production plans are a function of the demands' due dates and the items' earliness and tardiness penalties.

{\JITBP} is very challenging. It combines/extends two $\mathcal{NP}$-hard combinatorial optimization problems.  Its first component is a two-dimensional packing problem that has been extensively addressed via a panoply of techniques varying from OR to Artificial Intelligence (AI) to EC~ \cite{Lodi2014107,Burke2012,Polyakovsky2009767,Sim201537}.  Its second component is a just in time single machine batch scheduling problem~\cite{Hazır2012} that has been tackled via different EC approaches with the most prominent ones enumerated in~\cite{Hallah2015}.  It incorporates non-overlap, containment, assignment, disjunctive, and sequencing constraints whose simultaneous satisfaction is difficult. In addition, its search space, as for most packing related problems, contains a large number of infeasible and of symmetrical solutions. Optimizing {\JITBP} is further complicated by the non-traditional interdependence of its two components.  Relaxing the packing density (by assigning a single item per bin) does not reduce earliness-tardiness penalties while a dense production plan that cuts multiple items from a bin gives not only a tighter packing, but also smaller cutting times.  In fact, a faster production of the items supports the minimization of the earliness-tardiness.  In summary, determining the optimal number of bins is not trivial.

We address {\JITBP} via two approximate approaches: {\CP} and {\AB}.  {\CP} is a constraint programming (CP) search that explores constraint propagation. CP generates feasible solutions efficiently thanks to its flexible modelling framework, which exploits the structure of a model to direct and accelerate the search. Here, {\CP} adopts the impact-based search strategy, implemented in the general-purpose solver IBM CP Optimizer \cite{Refalo2004}.

{\AB}, on the other hand, is a hybrid constructive heuristic that searches the solution space via agent based modelling, identifies a feasible packing via CP techniques and optimally schedules the bins via linear programming.  It builds a solution through repeated negotiations between the set of agents representing the items and the set representing the bins. The agents cooperate to minimize the weighted earliness-tardiness penalties.    The choice of an agent-based modelling technique is motivated by (i) its success in application to bin packing \cite{Polyakovsky2009767} and just-in-time scheduling \cite{Polyakovskiy2014115} problems, (ii) the numerous infeasible and symmetric solutions encountered during the search, and (iii) the reported computational results, which confirm its efficiency in comparison to several enumerative techniques.

Sections \ref{sec:CPBA} and \ref{sec:AB} detail {\CP} and {\AB}. Section \ref{sec:Results} discusses the results. Section \ref{sec:Conclusion} is a summary.

\section{Constraint Programming Search}\label{sec:CPBA}
Because CP search techniques are  prominent for bin packing and scheduling problems, we formulate {\JITBP} as a CP model that combines both components and that we approximately solve with a general-purpose commercial solver. In CP, a problem is defined via a set of variables $\mathcal{X}$ and a set of constraints $\mathcal{C}$ given on $\mathcal{X}$ \cite{Bockmayr05,H02}. A variable $x_i \!\in\! \mathcal{X}$ can assume any value of its domain $D(x_i)$. To find an optimum, CP enumerates solutions subject to the constraint store $\mathcal{C}$ using a search tree where every variable $x_i \in \mathcal{X}$ is examined within some node. When $x_i$ is instantiated, the search inspects those constraints that share $x_i$.

In CP, each constraint is viewed as a special-purpose procedure that operates on a solution space. In fact, each procedure is a filtering algorithm that excludes, from the domains of the variables, those values that lead to infeasible solutions. CP relies on constraint propagation. Fixing the value of $x_i$ may eliminate some values from the domains of other variables that are connected to $x_i$ via one or more constraints of $\mathcal{C}$. That is, constraints sharing some common variables are linked to each other. Subsequently, the results of one filtering procedure are propagated to the others. CP calls the filtering algorithms repeatedly in order to achieve a certain level of consistency. When it yields at some stage $D(x_i) \!=\! \emptyset$, the filtering signals an infeasible solution (i.e., an inconsistent set of constraints). When $\left|D(x_i)\right|\!>\!1$, CP branches on $x_i$ by partitioning $D(x_i)$ into unique values, each corresponding to a branch. As the search descends into the tree, constraint propagation reduces the size of the domains of the variables. CP obtains a feasible solution when $|D(x_i)|=1$ for all $x_i$. When emphasis is on optimality, the search continues until either the optimum is found, or the exploration of the whole search tree is unsuccessful.

The performance of a CP model depends on its solver; specifically, on the filtering algorithms and on the search strategies it applies. Here, we resort to IBM ILOG \textsc{CP Optimizer} 12.6.2 with its searching algorithm set to the \textit{restart mode}. This mode adopts a general purpose search strategy ~\cite{Refalo2004} inspired from integer programming techniques and based on the concept of the impact of a variable. The impact measures the importance of a variable in reducing the search space. The impacts, which are learned from the observation of the domains' reduction during the search, help the restart mode dramatically improve the performance of the search.

\subsection{Decision Variables}\label{sec:DV}
\begin{figure}[tb]
\centering
\includegraphics[width=0.6\columnwidth]{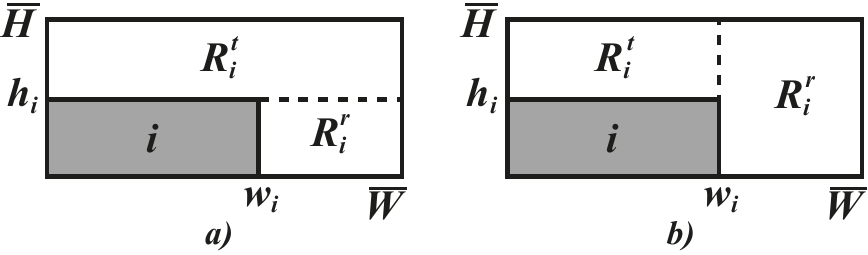}
\caption{Illustrating the free regions resulting from a horizontal (when variable $v_i=0$) and a vertical (when $v_i=1$) cut}
\label{fig1}
\end{figure}

The CP model for {\JITBP} explores the features of the problem. First, any item $i \!\in\! N$ can be obtained from a bin (or from any region of a bin) via a sequence of horizontal and vertical cuts, as Figure \ref{fig1} illustrates.  When the first applied cut is horizontal, as depicted in Figure \ref{fig1}.a, the bin $\textstyle\left(\overbar{W},\overbar{H}\right)$ is split into two regions: $R_i^t$ of size $\left(\overbar{W}, \overbar{H} \!-\! h_i\right)$ at the top of $i,$ and $R_i^r$ of size $\left(\overbar{W} \!-\! w_i, h_i \right)$ at the right of $i$. Hereafter, we refer to this case as a cutting pattern $a$. When the first cut is vertical, as shown in Figure \ref{fig1}.b, the cutting of the bin produces two regions: $R_i^t$ of size $\left(w_i, \overbar{H} \!-\! h_i\right)$ at the top of $i$ and $R_i^r$ of size $\left(\overbar{W} \!-\! w_i, \overbar{H} \right)$ at the right of $i$. Herein, we designate this pattern as $b$. Therefore, the extraction of any item always generates two regions: one to the top and one to the right of the item. Clearly, some of the regions may get a zero area if $\overbar{W} = w_i$ or $\overbar{H} = h_i.$

Suppose that the number of bins $m\!=\!n.$ The set $R$ of possible regions where the $n$ items may be positioned has $3n$ regions: the first $n$ regions emanate from the $n$ initial empty bins, the second $n$ regions correspond to $R_1^t,\ldots,R_n^t$, and the last $n$ regions to $R_1^r,\ldots,R_n^r$. In fact, the last $2n$ regions result from extracting the $n$ items. Thus, $R=\{R_1,\ldots,R_n,R_1^t,\ldots,R_n^t,R_1^r,\ldots,R_n^r\}$, where $R_1=\ldots=R_n=\left(\overbar{W},\overbar{H}\right)$. Evidently, this assumes that at most one item can be assigned to a region.
\setlength{\belowcaptionskip}{0.2\baselineskip plus 0.2\baselineskip minus 0.5\baselineskip}
Let $r=\left(r_1,\ldots,r_{3n}\right)$ be an integer-valued vector variable whose $j$th entry $r_j,\ j=1,\ldots,3n,$ is the item packed in the $j$th region; i.e., $r_j=i$ if region $j$ contains item $i$ and 0 if $j$ contains no item. Its domain is therefore
\formula{D\!\left(r_j\right) \!=\! \left\{0,\ldots,n\right\}, \, \fa{j}{R}}{cp:a1}
\noindent When $j \in \left\{1,\ldots,n\right\}$, item $r_j$ is the first item in bin $j.$ When $j\in\left\{n+1,\ldots,2n\right\},\ r_j$ is placed into $R_{j-n}^t$ on top of item $j-n.\ r_j \neq \left(j-n\right)$ since $R_{j-n}^t$ is the result of cutting off item $\left(j-n\right).$ Thus, the domain $D(r_j)$ excludes $j-n:$
\formula{r_j \!\neq\! j\!-\!n, \,\fa{j}{\left\{n\!+\!1,\ldots,2n\right\}}}{cp:a2}
\noindent Finally, when $j\in\left\{2n+1,\ldots,3n\right\},\ r_j$ is placed in $R_{j-2n}^r;$ that is, at the right of item $\left(j-2n\right)$ and $D(r_j)$ is free of element $j-2n:$
\formula{r_j \!\neq\! j\!-\!2n, \, \fa{j}{\left\{2n\!+\!1,\ldots,3n\right\}}}{cp:a3}
We compute $\lbfs,$ the lower bound of Fekete and Schepers~\cite{Fekete04} on the number of bins required to cut the $n$ items. We force the first $\lbfs$ regions to have exactly one item. Therefore the domain $D(r_j)$ of $r_j,\ j=1,\ldots,lb_{fs},$ excludes $\{0\}$:
\formula{r_j \!\neq\! 0, \, \fa{j}{\left\{1,\ldots,\lbfs\right\}}}{cp:a4}
This rule partially excludes symmetrical solutions arising when a filled bin $k$ empties all its items into an empty bin $k'$ that precedes or succeeds bin $k$ on the cutting machine. This exchange of items produces a different solution but the same objective function value. An empty bin has a zero processing time.

Similarly, let $e\!=\!\left(e_1,\ldots,e_n\right)$ be an integer-valued vector of variables reflecting the assignment of regions to items such that $e_i\!=\!j$ if item $i$ is positioned in region $j$. When $e_i \!\in\! \left\{1,\ldots,n\right\},\ i$ is the first packed item into its bin. On the other hand, when $e_i \!\in\! \left\{n+1,\ldots,2n\right\}$ or $e_i \in \left\{2n+1,\ldots,3n\right\},\ i$ is located in region $R_{e_i-n}^t$ at the top of item $\left(e_i-n\right)$ or in region $R_{e_i-2n}^r$ at the top of item $e_i-2n$, respectively. The domain of variable $e_i$ is
\formula{D\!\left(e_i\right) \!=\! \left\{1,\ldots,3n\right\}, \, \fa{i}{N}}{cp:a5}
\noindent Because item $i$ can not be packed into $R_{n+i}$ and $R_{2n+i}$, which are respectively $R_{i}^t$ and $R_{i}^r$, the following holds
\formula{\left(e_i \!\neq\! n\!+\!i \right) \!\wedge\! \left(e_i \!\neq\! 2n\!+\!i \right), \, \fa{i}{N}}{cp:a6}
Along with the aforementioned variables and constraints, the CP model uses an additional \textbf{five} sets of variables.

The \textbf{first} set has two integer-valued vectors $W \!\in\! \left\{\mathbb{N}_{\geq0}\right\}^{3n}$ and $H \!\in\! \left\{\mathbb{N}_{\geq0}\right\}^{3n}$  representing the widths and heights of the regions, where $\left(W_j,H_j\right),\ j\!=\!1,\ldots,3n,$ is the size of $R_j \!\in\! R$. Because $R_i\!=\!(\overbar{W},\overbar{H}),\ R_{n+i}\!=\!R_{i}^t,$ and $R_{2n+i}\!=\!R_{i}^r,$ for $i\!=\!1,\ldots,n,$
{\small
\begin{flalign}
& D\!\left(W_j\right) \!=\! \overbar{W},\, D\!\left(H_j\right) \!=\! \overbar{H} \!, \; \fa{j}{\left\{1,\ldots,n\right\}} \label{cp:a7}
\\
\nonumber &D\!\left(W_j\right) \!=\! \left\{w_{j-n},\dots,\overbar{W}\right\}\!,\; D\!\left(H_j\right) \!=\! \left\{0,\dots,\overbar{H}\!-\!h_{j-n}\right\}\!,
\\
& \qquad\qquad\qquad\qquad\qquad\qquad\qquad\qquad\qquad \fa{j}{\left\{n\!+\!1,\ldots,2n\right\}} \label{cp:a8}
\\
\nonumber& D\!\left(W_j\right) \!=\! \left\{0,\dots,\overbar{W}\!-\!w_{j-2n}\right\}\!,\; D\!\left(H_j\right) \!=\! \left\{h_{j-2n},\dots,\overbar{H}\right\}\!,
\\
& \qquad\qquad\qquad\qquad\qquad\qquad\qquad\qquad\qquad \fa{j}{\left\{2n\!+\!1,\ldots,3n\right\}} \label{cp:a9}
\end{flalign}
}
The \textbf{second} set has $x \!\in\! \left\{\mathbb{N}_{\geq0}\right\}^{3n}$ and $y \!\in\! \left\{\mathbb{N}_{\geq0}\right\}^{3n},$ two integer-valued vectors representing the bottom-left corner coordinates of the regions, where $\left(x_j,y_j\right),\ j=1,\ldots,3n,$ refers to the coordinates of $R_j \!\in\! R$. These variables abide to the conditions:
{\small
\begin{flalign}
& D\!\left(x_j\right) \!=\! 0,\; D\!\left( y_j\right) \!=\! 0, \; \fa{j}{\left\{1,\ldots,n\right\}} \label{cp:a10}
\\
\nonumber &D\!\left(x_j\right) \!=\! \left\{0,\dots,\overbar{W}-w_{j-n}\right\}\!,\; D\!\left(y_j\right) \!=\! \left\{h_{j-n},\dots,\overbar{H}\right\}\!,
\\
& \qquad\qquad\qquad\qquad\qquad\qquad\qquad\qquad\qquad  \fa{j}{\left\{n\!+\!1,\ldots,2n\right\}} \label{cp:a11}
\\
\nonumber& D\!\left(x_j\right) \!=\! \left\{w_{j-2n},\dots,W\right\}\!,\; D\!\left(x_j\right) \!=\! \left\{0,\dots,\overbar{H} \!-\! h_{j-2n}\right\}\!,
\\
& \qquad\qquad\qquad\qquad\qquad\qquad\qquad\qquad\qquad  \fa{j}{\left\{2n\!+\!1,\ldots,3n\right\}} \label{cp:a12}
\end{flalign}
}
The \textbf{third} set has an integer-valued vector $s \!\in\! \left\{1,\ldots,n\right\}^{3n}$ where $s_j \!=\! k$ when region $R_j \!\in\! R$ belongs to bin $k \!\in\! B$. The first $n$ regions correspond to the empty bins:
{\small
\begin{flalign}
&D\!\left(s_j\right) \!=\! \left\{1,\ldots,n\right\}, \, \fa{j}{\left\{1,\ldots,3n\right\}} \label{cp:a13}
\\[-0.1cm]
&s_j\!=\!j, \, \fa{j}{\left\{1,\ldots,n\right\}} \label{cp:a14}
\end{flalign}
}
The \textbf{fourth} set has $v \!\in\! \left\{0,1\right\}^{n}$ a binary vector defining the cutting pattern used to extract the item: $v_i=0$ if the first cut that yields item $i$ is horizontal (cf. Figure \ref{fig1}.a), and 1 otherwise. Thus,
\formula{D\!\left(v_i\right) \!=\! \left\{0,1\right\}, \, \fa{i}{N}}{cp:a15}
The \textbf{fifth} set has $t\!\in\! \left\{\mathbb{N}_{\geq 0}\right\}^{n}$ and $c \!\in\! \left\{\mathbb{N}_{\geq 0}\right\}^{n},$ two integer-valued vectors that define the processing times and the completion times of the bins. Let $t_{max}$ and $c_{max}$ be upper bounds on the maximal processing and completion time of a bin, respectively. Then
{\small
\begin{flalign}
&D\!\left(t_k\right) \!=\! \left\{0,\ldots,t_{max}\right\}, \, \fa{k}{B} \label{cp:a16}
\\[-0.1cm]
&D\!\left(c_k\right) \!=\! \left\{0,\ldots,c_{max}\right\}, \, \fa{k}{B} \label{cp:a17}
\end{flalign}
}
\vspace{-.4cm}
%
\subsection{CP Model}\label{sec:MODEL}
The CP model uses constraint $\ad{g_1}{g_n},$ which ensures that the variables of $\left[g_1,\ldots,g_n\right]$ take distinct values. Additionally, the model uses three expressions: $\eq{g}{h},$ which returns $1$ or $true$ if $g=h$ and $0$ or $false$ otherwise; $\el{g}{h},$ which returns the $h$th variable in the list of variables $g$; and $\countn{g}{h},$ which counts the number of variables of $g$ taking the value $h$.
The CP model follows.
{\footnotesize
\begin{flalign}
\nonumber \mbox{min} \,\,& \textstyle\sum_{i \in N} \!\textstyle\sum_{k \in B} \eq{{\el{s}{e_i}}}{k} \cdot
\\[-0.05cm]
& \left(\epsilon_i \!\!\cdot\! \max\!\left\{0,\!d_i \!-\! c_k\right\} \!+\! \tau_i \!\!\cdot\! \max\!\left\{0,\!c_k \!-\! d_i\right\} \right) \label{cpC:0}
\\[-0.05cm]
\nonumber \mbox{s.t.} \,\,& \text{(\ref{cp:a1})-(\ref{cp:a17})}
\\[-0.05cm]
& \el{r}{e_i} \!=\! i &  \fa{i}{N} \label{cpC:1}
\\[-0.05cm]
& \ad{e_1}{e_n} \label{cpC:2}
\\[-0.05cm]
& \countn{r}{i}\!=\!1 &  \fa{i}{N} \label{cpC:15}
\\[-0.05cm]
& w_i \!\le\!\el{W}{e_i} &  \fa{i}{N} \label{cpC:3}
\\[-0.05cm]
& h_i \!\le\!\el{H}{e_i} &  \fa{i}{N} \label{cpC:4}
\\[-0.05cm]
& W_{n+i} \!\le\!\el{W}{e_i} \cdot \left(1 \!-\! v_i\right) \!+\! w_iv_i & \fa{i}{N} \label{cpC:5}
\\[-0.05cm]
& H_{n+i} \!\le\!\el{H}{e_i} \!-\! h_i &  \fa{i}{N} \label{cpC:6}
\\[-0.05cm]
& W_{2n+i} \!\le\!\el{W}{e_i} \!-\! w_i &  \fa{i}{N} \label{cpC:7}
\\[-0.05cm]
& H_{2n+i} \!\le\!h_i\left(1 \!-\! v_i\right) \!+\! \el{H}{e_i} \cdot v_i &  \fa{i}{N} \label{cpC:8}
\\[-0.05cm]
& x_{n+i} \!=\! \el{x}{e_i} &  \fa{i}{N} \label{cpC:9}
\\[-0.05cm]
& y_{n+i} \!=\! \el{y}{e_i}\!+\!h_i &  \fa{i}{N} \label{cpC:10}
\\[-0.05cm]
& x_{2n+i} \!=\! \el{x}{e_i}\!+\!w_i &  \fa{i}{N} \label{cpC:11}
\\[-0.05cm]
& y_{2n+i} \!=\! \el{y}{e_i} &  \fa{i}{N} \label{cpC:12}
\\[-0.05cm]
& s_{n+i} \!=\! \el{s}{e_i} &  \fa{i}{N} \label{cpC:13}
\\[-0.05cm]
& s_{2n+i} \!=\! \el{s}{e_i} &  \fa{i}{N} \label{cpC:14}
\\[-0.05cm]
&  \eq{r_k}{0} \rightarrow \!\!\textstyle\sum_{l=k+1}^n {r_l} \!=\! 0 &  k \!\!\in\!\! \left\{\lbfs\!\!+\!\!1,\!...,\!n\right\} \label{cpC:16}
\\
& \scalebox{.98}[1.0]{$t_k \!=\! f \!\left( \cup_{i=1}^n { \left\{e_i \!\!:\! \eq{{\el{s}{e_i}}}{k} \right\}}\right)\!$} & \fa{k}{B} \label{cpC:17}
\\[-0.05cm]
& c_k \leq c_{k+1} \!-\! p_{k+1} & k \!\!\in\! B \! \setminus \!\! \left\{m \right\} \label{cpC:18}
\\[-0.05cm]
& c_1 \geq p_1 & \label{cpC:19}
\end{flalign}
}

Expression (\ref{cpC:0}) calculates the objective value as the weighted sum of earliness-tardiness penalties. Because $e_i$ is the region to which item $i$ is assigned, $\el{s}{e_i}$ in Eq. (\ref{cpC:0}) gives the index of the bin where $i$ is packed. When $\eq{{\el{s}{e_i}}}{k}$ confirms that $i$ is assigned to bin $k,$ the expression computes the weighted earliness tardiness of $i$ based on $c_k$. Constraint (\ref{cpC:1}) establishes a dual notation by forcing region $r_{e_i}$ to contain item $i$ when $i$ is assigned to $e_i$. Constraint (\ref{cpC:2}) verifies that all the items are assigned to different regions. Constraint (\ref{cpC:15}) ensures that exactly one region contains item $i$. Constraints (\ref{cpC:3}) and (\ref{cpC:4}) guarantee that $e_i,$ which holds item $i,$ is large enough to fit $i$. Constraints (\ref{cpC:5}) and (\ref{cpC:6}) bound the size of the top residual area $R_i^t$ obtained when cutting item $i$. When the first cut generating $i$ is horizontal (i.e., $v_i\!=\!0$), Eq. (\ref{cpC:5}) limits the width of $R_i^t$ to the width of region $e_i$ where $i$ is positioned. On the other hand, when the first cut generating $i$ is vertical (i.e., $v_i\!=\!1$), Eq. (\ref{cpC:5}) limits the width of $R_i^t$ to the width $w_i$. Constraint (\ref{cpC:6}) bounds the height of $R_i^t$ by the difference between the height of $e_i$ and $h_i$. Indeed, it does not depend on the cutting pattern. Similarly, constraints (\ref{cpC:7}) and (\ref{cpC:8}) bound the size of $R_i^r$ with Constraint (\ref{cpC:7}) delimiting the width of $R_i^r$ by the difference between the width of $e_i$ and $w_i$ and Constraint (\ref{cpC:8}) limiting the height of $R_i^r$  to $h_i$ when $i$ is extracted with respect to pattern $a$ (i.e., $v_i=0$) and to the height of $e_i$ when $i$ is positioned and cut according to pattern $b$.  Constraint (\ref{cpC:9}) sets the $x$-coordinate of region $R_i^t$ to $e_i$'s $x$-coordinate. Constraint (\ref{cpC:10}) computes the $y$-coordinate of region $R_i^t$ as the sum of the $y$-coordinate of region $e_i$ and $h_i$. Constraint (\ref{cpC:11}) computes the $x$-coordinate of region $R_i^r$ as the sum of the $x$-coordinate of region $e_i$ and $x_i$. Constraint (\ref{cpC:12}) sets the $y$-coordinate of region $R_i^r$ to the $y$-coordinate of region $e_i$. Constraints (\ref{cpC:13}) and (\ref{cpC:14}) define, respectively, the bin to which regions $R_i^t$ and $R_i^r$ belong. Constraint (\ref{cpC:16}) implies that no bin succeeding bin $k$ can hold a region if $k$ is empty. This is a symmetry breaking constraint that prohibits filling a bin $\left(k\!+\!1\right)$ if bin $k$ is not filled. Constraint (\ref{cpC:17}) calculates the processing time of bin $k$. Constraint (\ref{cpC:18}) guarantees the no overlap of the processing windows of two consecutive bins. Finally, constraint (\ref{cpC:19}) ensures that the schedule starts after time zero.  This CP model solves {\JITBP} exactly even with the restart mode. However, {\CP} uses it to find an approximate solution as it presets its runtime.

Within the search, items are assigned to regions in the ascending order of their due dates. The solver instantiates $e_1,\ldots,e_n;$ then applies its default strategy to the other variables.  Our extensive study shows that such an order gives the best results fast.

\section{Agent-Based Heuristic}\label{sec:AB}
{\AB} is a constructive heuristic that packs items into bins through negotiation, and schedules the packed bins optimally using a linear program.  It uses a packing procedure {\pack} that searches for a feasible guillotine packing of a set of items into a single bin via a reduced version of {\CP}. Because it solves a feasibility problem, {\pack} omits Eq. ({\ref{cpC:0}}) and drops the $s$, $t$ and $c$ variables along with their related constraints, i.e. Eqs. (\ref{cp:a13}-\ref{cp:a14},\ref{cp:a16}-\ref{cp:a17},\ref{cpC:13}-\ref{cpC:19}). {\pack} is solved via IBM ILOG's CP Optimizer, which is allocated a preset threshold runtime.  It either returns a feasible packing or signals the infeasibility of the grouping of the items in a bin.

Suppose that $m$ is fixed and that an integer-valued vector $x\!\in\!\left\{\mathbb{N}_{\geq0}\right\}^n$ represents an assignment of items to bins such that $x_i=k$ if item $i \!\in\! N_k$. For a given $x,$ the processing time of bin $k \!\in\! B$ is known. Thus, the completion times that minimize the weighted earliness tardiness of the items is the optimal solution to the linear program $\OBJ{N}$ whose decision variables are: $c_k\geq0,\ T_i\geq 0$ and $E_i\geq 0$ for $k \!\in\! B$ and $i \!\in\! N$. $\OBJ{N}$ follows.
{\small
\begin{flalign}
\mbox{min} \,\,& \textstyle\sum_{i \in N} \left(\epsilon_i E_i + \tau_i T_i\right) & \label{eq:1}
\\[-0.1cm]
\nonumber \mbox{s.t.} \,\,& \text{(\ref{cpC:18})-(\ref{cpC:19})} &\\[-0.1cm]
& T_i-E_i = c_k - d_i & \fa{i}{N},\; \fa{k}{B\,:\, x_i=k} \label{eq:2}
\\[-0.1cm]
& T_i,\;E_i\in \mathbb{R}_{\geq0} & \fa{i}{N} \label{eq:3}
\\[-0.1cm]
& c_k\in \mathbb{R}_{\geq0} & \fa{k}{B} \label{eq:4}
\end{flalign}
}
$\OBJ{N}$ schedules the bins on the single cutter and inserts idle time between successive bins if this decreases the total weighted earliness tardiness, defined by Eq. (\ref{eq:1}). Eq. (\ref{cpC:18}) determines the completion time of each bin ensuring that the processing periods of two successive bins do not overlap in time. Eq. (\ref{cpC:19}) guarantees that the schedule starts after time zero. Eq. (\ref{eq:2}) calculates the tardiness $T_i$ and the earliness $E_i$ of job $i$ when $i$ is assigned to bin $k$. Finally, Eqs. (\ref{eq:3}-\ref{eq:4}) declare the variables positive. $\OBJ{N}$ is a linear program that can be solved via IBM ILOG's CPLEX.

\subsection{Initial Partial Solution}\label{sec:ps}
{\AB} starts with constructing a partial solution. \textbf{First}, {\AB} computes the lower bound $\lbfs$ of Fekete and Schepers~\cite{Fekete04} on the number of bins required to pack the $n$ items. Then {\AB} constructs a partial solution with $m\!=\!\lbfs$ empty bins.

\textbf{Second}, {\AB} uses a peak clustering algorithm \cite{Yager} that identifies items with close due dates. These items constitute a bottleneck on the machine. The idea is that a bin groups a set of items such that its completion time lies at the centre of the due dates of the items. Assigning items causing a bottleneck to the same bin reduces the lateness of the items; in particular when the items' per unit earliness tardiness penalties are symmetric. The clustering algorithm finds the $k$th cluster by evaluating, for $i \!\in\! N,$ a clustering function
$$\phi_i^{k} =
\begin{cases}
\textstyle\sum_{i' \in N} \exp \left(- \frac{{\lvert d_i - d_{i'} \rvert}^2}{(.5r)^2 }\right), & \mbox{if } k=1 \\
\phi_i^{k-1} - \phi_{i_{k-1}^*}^{k-1} \exp \left(- \frac{{\lvert d_i - d_{i_{k-1}^*} \rvert}^2}{(.5r)^2} \right), & \mbox{if } k>1
\end{cases}$$
\noindent where $r$ is the parameter that governs the width of the peak function. It then selects $i_k^*,$ the item with the largest peak function value: $\phi_{i_k^*}^k = \max_{i \in N} \{{\phi_i^k}\}$. It removes $i_k^*$ from $N$ and places it in a list $\overbar{U}$ of highly bottleneck items. It updates the modified peak function and iterates the process until it has removed $m$ bottleneck items. The $n-m$ unpacked items constitute the set $U$.

\subsection{Completing a Partial Solution}\label{sec:CS}
Let $\WETbin{i^*}{k}=$
\begin{center}
$\displaystyle \sum_{i\in\overbar{U}\cup\left\{i^*\right\}} \sum_{k' \in B:x_i=k'}  \epsilon_i \!\cdot\! \max\left\{0,d_i \!-\! c_{k'}\right\} \!+\! \tau_i \!\cdot\! \max\left\{0,c_{k'} \!-\! d_i\right\}$
\end{center}
denote the objective value of the current partial solution augmented by $\{i^*\}$ were $i^*$ to leave $U$ and be packed in $k$. $\WETbin{i^*}{k}$ is the value of the optimum of $\OBJ{\overbar{U}\!\cup\! \{i^*\}}$.

The items in $U$ and the filled bins act as individual greedy agents that undertake negotiation actions. A bin $k$ seeks to attract the best item $i^*\!\in\! U$, defined as the item which minimises $\WETbin{i^*}{k}$. Likewise, item $i \!\in\! U$ seeks the best bin $k^*$ where $k^*$ minimises $\WETbin{i}{k^*}$ should $i$ leave $U$ and join $k^*$. The item-attraction and bin-seeking negotiation actions, named as {\GroupFormation} and {\GroupJoin} respectively, continue until all items are assigned to bins $\left(U \!=\! \emptyset \right)$ or none of the remaining items of $U$ are packable into existing bins. At this point, a repacking procedure is triggered.

Every time a bin's {\GroupFormation} action completes successfully, {\AB} runs a local search that obtains an optimised partial solution before {\AB} undertakes its next assignment decision.  Differently stated, the local search downsizes the myopia of the greedy actions of the agents and keeps the partial solution balanced during the construction phase.  It employs two operators: \textsl{insertion}, which moves an item from bin $k$ to its neighbouring bin $k\!+\!1$ or vice versa, and  \textsl{swap}, which exchanges two items, $i \!\in\! N_k$ and $i'\!\in\! N_{k+1},$ from two neighbouring bins. A move or exchange is adopted when it improves the current solution and maintains its feasibility.  The local search applies the operators sequentially until it obtains a better solution.  It operates on neighboring bins only in order to have a short run time.  However, it may consider a larger neighbourhood when the earliness and tardiness costs are highly asymmetric.

\subsubsection{\textbf{GroupFormation}}\label{sec:GF}
The {\GroupFormation} action, detailed in Algorithm \ref{Alg:GF}, considers the items in $U$ in ascending order of their $\textstyle \epsilon_i E_i + \tau_i T_i,$ where $E_i$ and $T_i$ are computed with respect to the current $c_{k}$ of bin $k$. Evidently, the current $c_{k}$ is an estimate of the true $c_{k},$ which changes as items of $U$ are assigned to $k$. For each of the first $\rho$ items of $U$, the {\GroupFormation} action of bin $k$ computes, using $\OBJ{N}$, the true $c_{k}$ were this item to join $k$. It considers the $\rho$ most prominent items first to reduce its run time.

Next, the {\GroupFormation} action of a bin $k$ chooses the candidate $i^*$ which induces the smallest $\WETbin{i^*}{k}$ were $i^*$ to join $k$. It checks whether $k$ can append $i^*$. If the $\lbfs$ lower bound for $N_{k}\!\cup\!\left\{i^*\right\}$, denoted as $\lbfs\left(N_{k}\!\cup\!\left\{i^*\right\} \right),$ is less than or equal to 1, $i^*$ may be packed in $k$. Therefore, the {\GroupFormation} action triggers {\pack}, which searches for a feasible guillotine packing of $i^*$ and the items already assigned to $k$.  When {\pack} returns a valid packing, $k$ sends a join request to $i^*$, which has independently undertaken its own {\GroupJoin} action. When $i^*$ accepts the request, the {\GroupFormation} action assigns $i^*$ to $k$, and terminates successfully.
{\pack} may fail to find a further packing $i^*$ into $k$ either because it runs out of time or it has exhausted all possible packing possibilities. In this case, the {\GroupFormation} action of $k$ removes $i^*$ from the list of candidates items and selects the next item, say $i^{**},$ that minimizes $\WETbin{i^{**}}{k}$ were $i^{**}$ to join $k$. When $k$ fails to append any of the $\rho$ candidate items, it iteratively considers the next item of $U$, compute the true $c_{k}$ were this item to join $k$ and checks the feasibility of packing this item into $k$. If it still fails to append any of the items of $U$, the {\GroupFormation} action terminates unsuccessfully. When all available bins fail their {\GroupFormation} actions while $U \!\neq\!\emptyset,$ the repacking procedure of Section~\ref{sec:Repacking} is applied.
\begin{algorithm}[t]
{\small
\caption{{\GroupFormation} action of bin $k$}\label{Alg:GF}
\begin{adjustwidth}{-0.3cm}{}
\begin{algorithmic}
\State Sort $U$ in the ascending order of $\epsilon_i E_i \!+\! \tau_i T_i$, $i\!\!\in\!\!U$, with respect to $c_{k}$;
\State Initialise the list of candidates $\Gamma=\emptyset$ and the list of examined items $\Delta=\emptyset$;
\State Select the first $\rho$ items in $U$ and add them to $\Gamma$;
\While{($true$)}
	\For{($\forall i \in \Gamma$)}
		\If {($\alpha_{ki}=-1$)} $\alpha_{ki}$ = $\WETbin{i}{k}$;
		\EndIf
	\EndFor
	\State Select $i^* = \min_{i \in \Gamma}\{\alpha_{ki}\}$
	\If{($\lbfs\left(N_{k}\!\cup\!\left\{i^*\right\}\right)\leq 1$)}
		\If{($\lambda_{ki^*} = -1$)} $\lambda_{ki^*} = \text{\pack}\left(N_k\!\cup\!\left\{i^*\right\}\right)$; \EndIf
		\If{($\lambda_{ki^*} = 1$)}
			\If{(the {\GroupJoin} action of $i^*$ returns $true$)}
				\State Append $i^*$ to $N_k$ and remove it from $U$;
				\For{($\forall\alpha_{k'i} \in A:\,k'\in B,\,i\in N$)} Set $\alpha_{k'i}=-1$; \EndFor
				\For{($\forall\lambda_{ki} \in \Lambda:\,i\in N$)} Set $\lambda_{ki}=-1$; \EndFor
				\State Return $true;$
			\EndIf
		\EndIf
	\EndIf
	\State Add $i^*$ to $\Delta$ and remove it from $\Gamma$;
	\If{($\Gamma = \emptyset$)}
		\If{($\Delta = U$)} Return $false$; \EndIf
		\State Set $\Gamma = U \setminus \Delta$;
	\EndIf
\EndWhile
\end{algorithmic}
\end{adjustwidth}
}
\end{algorithm}
%
\subsubsection{\textbf{GroupJoin}}\label{sec:GJ}
The bin and item agents act individually and greedily. This can result in low-penalty items being attached to bins whose completion times are far from the items' due dates. This might also be problematic when the per unit penalties of the items assigned to a bin are highly asymmetric. It can quickly lead to a ripple effect, as these low-penalty items may force items with higher penalties not to be in their most proximal bins due to packing constraints. This, in turn, forces more items to be packed into less proximal bins; yielding a large increase of the penalties and an overall degeneration in solution quality. To avoid such a scenario, an item $i$ declines an attachment offer from bin $k$ if it can be packed into bin $k'$ and the overlap of $[d_{i}-p_{i},d_{i}]$ with $[c_{k'}-p_{k'},c_{k'}]$ is larger than the overlap of $[d_{i}-p_{i},d_{i}]$ with $[c_{k}-p_{k},c_{k}],$ where $[d_{i}-p_{i},d_{i}]$ is the ideal processing time of $i$ and $p_{i}$ is the processing time of a bin containing only item $i$.

Each item $i \!\in\! U$ is interested in joining a bin whose processing time window overlaps its own ideal processing window. Let $\mathcal{O}_i$ denote the set of such bins. The {\GroupJoin} action sorts the bins of $\mathcal{O}_i$ in descending order of the overlap, and chooses the bin $k' \!\in\! \mathcal{O}_i$ with the largest overlap. If $k'$ is the bin that sent $i$ the attachment offer, $i$ accepts the joining request. Otherwise, the {\GroupJoin} action checks whether $i$ can be packed in $k'$. If $\lbfs\left(N_{k'}\!\cup\!\left\{i\right\} \right)\! \leq\! 1$, it calls {\pack}. If the packing of $i$ in $k'$ is feasible, $i$ declines the invitation of $k$; otherwise, the {\GroupJoin} action considers the next bin of $\mathcal{O}_{i}.$ When the {\GroupJoin} action exhausts all bins of $\mathcal{O}_{i}$ without packing $i$, it proceeds to the following alternative plan. For each bin $k' \!\in\! B \!\setminus\! \mathcal{O}_i,$ it computes the penalty $\WETbin{i}{k'}$ were $i$ to join $k'$. It first forms the set of bins $B'$ such that $WET(i,k') \leq WET(i,k)$ and sorts it in ascending order of $WET(i,k').$ Then, it iteratively scans $B' \!\subseteq\! B \!\setminus\! \mathcal{O}_i$. If $k\!=\!k'$, $k'\!\in\! B'$, $i$ accepts the invitation of $k$ and the {\GroupJoin} action ends successfully. Otherwise, it tests whether $i$ is packable in $k'$ with a lower earliness tardiness penalty. If $\lbfs\left(N_{k'}\!\cup\!\left\{i\right\} \right)\! \leq\! 1$ and {\pack} returns a feasible packing, the {\GroupJoin} action declines the invitation of $k$ whereas the {\GroupJoin} action checks the next bin of $B'$ when packing is impossible. The {\GroupJoin} action ends successfully when it checks all the bins of $B'$ and does not find a bin proposing a placement yielding smaller earliness-tardiness penalties.  Algorithm \ref{ALG:JR} details the {\GroupJoin} action.
\begin{algorithm}[t]
{\small
\caption{{\GroupJoin} action of item $i$ when a join request from bin $k$ is received}\label{ALG:JR}
\begin{adjustwidth}{-0.3cm}{}
\begin{algorithmic}
\State Initialise the list of bins $\mathcal{O}_i\!=\!\cup_{k'\in B} \left\{k': [d_i\!-\!p_i,d_i] \!\cap\! [c_{k'}\!-\!p_{k'},c_{k'}] \!\neq \!\varnothing\! \right\}$;
\State Sort $\mathcal{O}_i$ in the descending order of the overlap $[c_{k'}-p_{k'},c_{k'}] \cap [d_i-p_i,d_i]$;
\For {($\forall k' \in \mathcal{O}_i$)}
	\If{($k'=k$)} Return $true$ \EndIf
	\If{($\lbfs\left(N_{k'}\!\cup\!\left\{i\right\}\right)\leq 1$)}
		\If{($\lambda_{k'i} = -1$)} Set $\lambda_{k'i} = \text{\pack}\left(N_{k'}\!\cup\!\left\{i\right\}\right)$; \EndIf
		\If{($\lambda_{k'i} = 1$)} Return $false$; \EndIf
	\EndIf
\EndFor
\For{($\forall k' \in B \setminus \mathcal{O}_i$)}
		\If {($\alpha_{k'i}=-1$)} $\alpha_{k'i}=\WETbin{i}{k'}$; \EndIf
\EndFor		
\State Form the list of bins $B'\!=\cup_{k'\in B \setminus \mathcal{O}_i} \left\{k': \WETbin{i}{k'}\leq\WETbin{i}{k}\right\}$;
\State Sort $B'$ in ascending order of $\WETbin{i}{k'}$, $k' \in B'$;
\For{($\forall k' \in B'$)}
	\If {($k'=k$)} Return $true$; \EndIf
	\If{($\lbfs\left(N_{k'}\!\cup\!\left\{i\right\}\right)\leq 1$)}
		\If {($\lambda_{k'i} = -1$)} $\lambda_{k'i}= \text{\pack}\left(N_{k'}\!\cup\!\left\{i\right\}\right)$; \EndIf
		\If {($\lambda_{k'i} = 1$)} Return $false$; \EndIf
	\EndIf
\EndFor
\State Return $true$;
\end{algorithmic}
\end{adjustwidth}
}
\end{algorithm}
%
\subsubsection {\textbf{Repacking}}\label{sec:Repacking}
When there remain unpacked items (i.e., $U\neq \emptyset$) while the {\GroupFormation} actions of all the bins fail, {\AB} triggers a repacking procedure. It identifies a bin $k^*$ whose completion time $c_{k^*}$ is the closest to the due date of the first unpacked item $i\!\in\!U$. It empties bin $k^*$ along with bins $(k^*-1)$ and $(k^*+1)$ (if they exist), and inserts their unpacked items into set $U$. It inserts a new bin immediately following bin $k$, increments the number of bins by one, and resumes the {\GroupFormation} actions.

\subsubsection {\textbf{Caching}}\label{sec:Caching}
{\AB} reduces its runtime by nearly an order of magnitude by caching results from previous objective function calls and packing feasibility checks. Consider bin $k \in B$ sending a join request to item $i \in U$. {\AB} doesn't recompute the total scheduling penalty resulting from $i$ joining $k$ if this penalty was computed previously and the schedule has since been unchanged. Similarly, {\AB} doesn't rerun {\pack} for $N_k \cup \left\{i\right\}$ if it was obtained at a prior stage and the content of the bin $k$ has since been unchanged. {\AB} proceeds similarly for the {\GroupJoin} actions when an item requests to join a bin.

Matrix $\Lambda \!\in\! \left\{-1,0,1\right\}^{m \times n}$ stores packing feasibility check results such that its element $\lambda_{ki},\ k\in B,\ i \in N,$ equals 0 if the packing of $i$ into $k$ is infeasible, equals 1 if it is feasible, and -1 if the result is unknown. Every time an item is added to a bin $k \in B$, all the elements of the row associated with $k$ are changed to the default unknown packing state; i.e., to $-1$.

Penalty computation results are stored in matrix $A \in \mathbb{R}_{\geq-1}^{m \times n}$ whose element $\alpha_{ki},\ k\in B,\  i \in N,$ equals the total penalty generated by attaching $i$ to $k$. Unlike the feasibility checks, every entry of $A$ is reset to the unknown state $-1$ when any bin adds a new item as this addition may alter the total scheduling penalty.

\section{Computational Results}\label{sec:Results}

{\CP} and {\AB} are implemented in C\#, with CP models solved via IBM's ILOG Optimisation Studio's CP solver version 12.6.2. They are run on a personal computer with 4GB of RAM and a 3.06GHz Dual Core processor.  Their performances are compared on instances that are randomly generated as follows:
\begin{itemize}[noitemsep, nosep, label=$\bullet$, leftmargin=*]
\item $n=$20, 40, 60, 80, and 100;
\item $\epsilon_i$ and $\tau_i$ following the uniform$[1,5]$;
\item the processing time function of non-empty bin $k$ is $f\left(N_k\right) = t^l + t^h\cdot\left|N_k\right| + t^c \cdot \textstyle\sum_{i\in N_k}\left(w_i\!+\!h_i\right)$, where bin loading time $t^l=100$, the item's handling time $t^h=30$, and the cutting time $t^c=0.02$;
\item items' due dates distributed according to: a normal with mean $\lambda$ and standard deviation $0.1\lambda,$ or a uniform[0, $2\lambda$] where $\lambda=\frac{t^l \sum_{i \in N} h_i w_i}{2 \overbar{W} \overbar{H}}$; and
\item classes of instance as given in Table \ref{tab:type}. They are those of     the most known benchmark set for two-dimensional bin packing \cite{Lodi2002379}. Classes 7-10 involve four types of items whose dimensions follow uniform distributions whose ranges are respectively:
{\small
\begin{itemize}[noitemsep, nosep, label=$-$, leftmargin=*]
	\item type 1: $\left(\left[\frac{2}{3}\overbar{W},\!\overbar{W}\right]\!,\!\left[1,\!\frac{1}{2}\overbar{H}\right]\right)$;
	\item type 2: $\left(\left[1,\!\frac{1}{2}\overbar{W}\right]\!,\!\left[\frac{2}{3}\overbar{H},\!\overbar{H}\right]\right)$;
	\item type 3: $\left(\left[\frac{1}{2}\overbar{W},\!\overbar{W}\right]\!,\!\left[\frac{1}{2}\overbar{H},\!\overbar{H}\right]\right)$; and
	\item type 4: $\left(\left[1,\!\frac{1}{2}\overbar{W}\right]\!,\!\left[1,\!\frac{1}{2}\overbar{H}\right]\right)$.
\end{itemize}
}
\end{itemize}
The instances are grouped according to their bin to item size ratios, which reflect the average number of items that can be packed into a bin:  A low bin density set $\mathcal{L}$ with relatively large items and a high bin density set $\mathcal{S}$ with small items.  As in real life furniture manufacturing, a bin's processing time accounts for its size and that of its items. For each class, problem size, and type of due date distribution, 10 instances are generated. The resulting 1000 problems, which will serve as benchmark instances for future research\footnote{The full set of test instances is available at http://cs.adelaide.edu.au/$\sim$optlog/}.
\begin{table}[tb]
\centering
\caption{Generation of the widths and heights of items}\label{tab:type}
{\scriptsize
\begin{tabularx}{\columnwidth}
{lXlp{5.5cm}}
\hline
Density&Class&\begin{color}{black}$\overbar{W}=\overbar{H}$\end{color}& $\left(w_i,h_i)\right)$\\
\hline
$\mathcal{L}$&1&10&uniform$\left[1,10\right]$\\
$\mathcal{S}$&2&30&uniform$\left[1,10\right]$\\
$\mathcal{L}$&3&40&uniform$\left[1,35\right]$\\
$\mathcal{S}$&4&100&uniform$\left[1,35\right]$\\
$\mathcal{L}$&5&100&uniform$\left[1,100\right]$\\
$\mathcal{S}$&6&300&uniform$\left[1,100\right]$\\
$\mathcal{L}$&7&100&type 1 with probability 70\%; type 2, 3, 4 with probability 10\% each\\
$\mathcal{L}$&8&100&type 2 with probability 70\%; type 1, 3, 4 with probability 10\% each\\
$\mathcal{L}$&9&100&type 3 with probability 70\%; type 1, 2, 4 with probability 10\% each\\
$\mathcal{S}$&10&100&type 4 with probability 70\%; type 1, 2, 3 with probability 10\% each\\
\hline
\end{tabularx}
}
\end{table}
{\AB} is run with $r\!=\!40$ and $\rho\!=\!3$. Both values are set as a result of extensive experiments. $r$ is not very influential but instance-dependent. Setting $r$ should avoid both very large and very small values. When $r$ is too large, the peak function accounts for too many due dates; consequently, the selection of bottlenecks becomes problematic as the clustering algorithm will tend to detect a single cluster rather than bottlenecks.  Similarly, when $r$ is too small, the clustering algorithm identifies too many bottlenecks; consequently, finding the bottleneck with the largest impact is hard. On the other hand, the value of $\rho$ is not critical as it slightly speeds up the {\GroupFormation} computations.  {\AB} allocates a 0.5 s runtime to the constraint program {\pack} that checks for the feasibility of a packing. All statistical inferences are valid at a 5\% significance level.
\begin{figure}
  \centering
  \includegraphics[width=0.8\columnwidth]{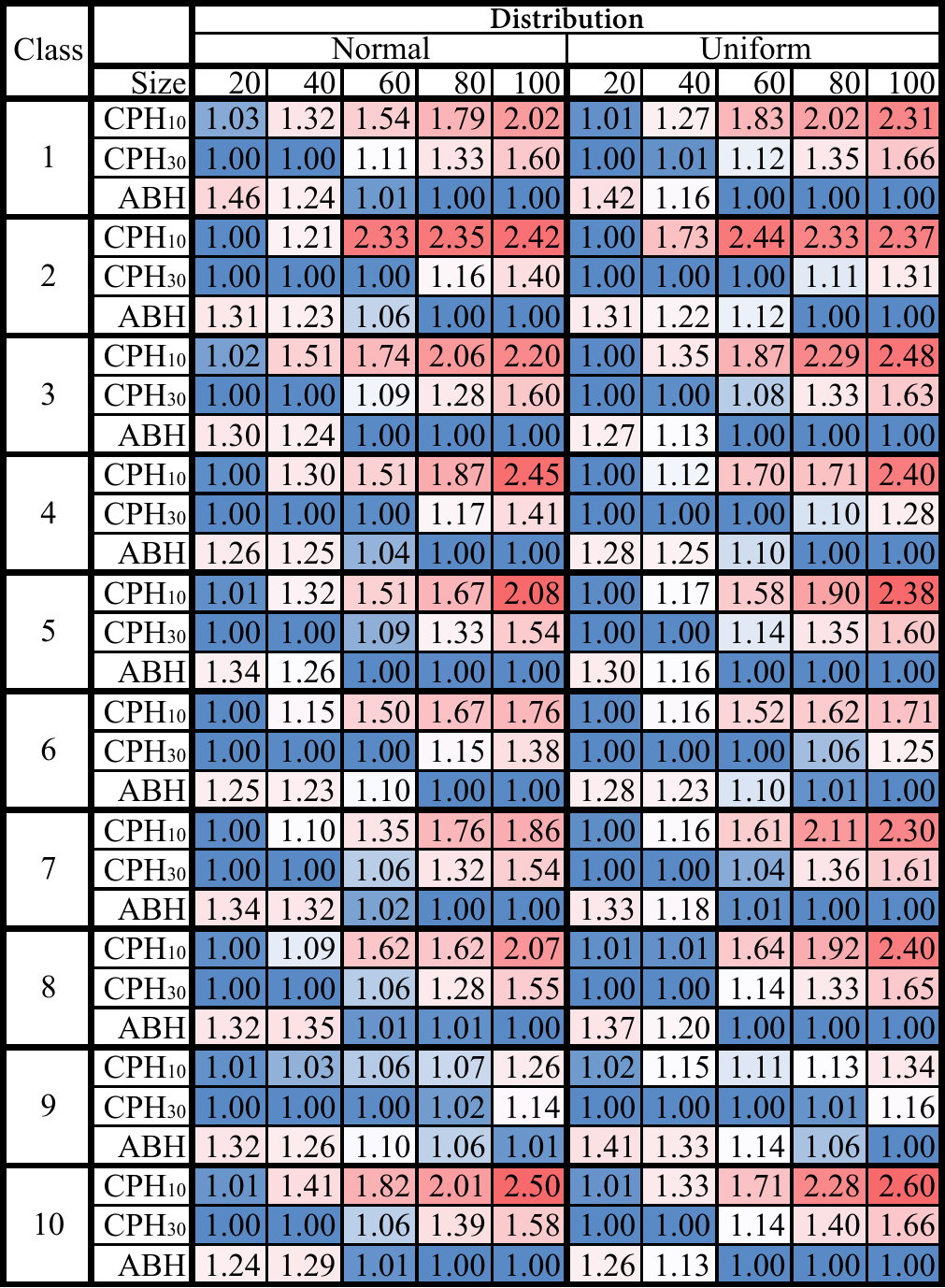}\\
  \caption{Heat map of performance ratio}\label{fig:results}
\end{figure}

\begin{figure}
\centering
 \includegraphics[width=0.6\columnwidth,keepaspectratio]{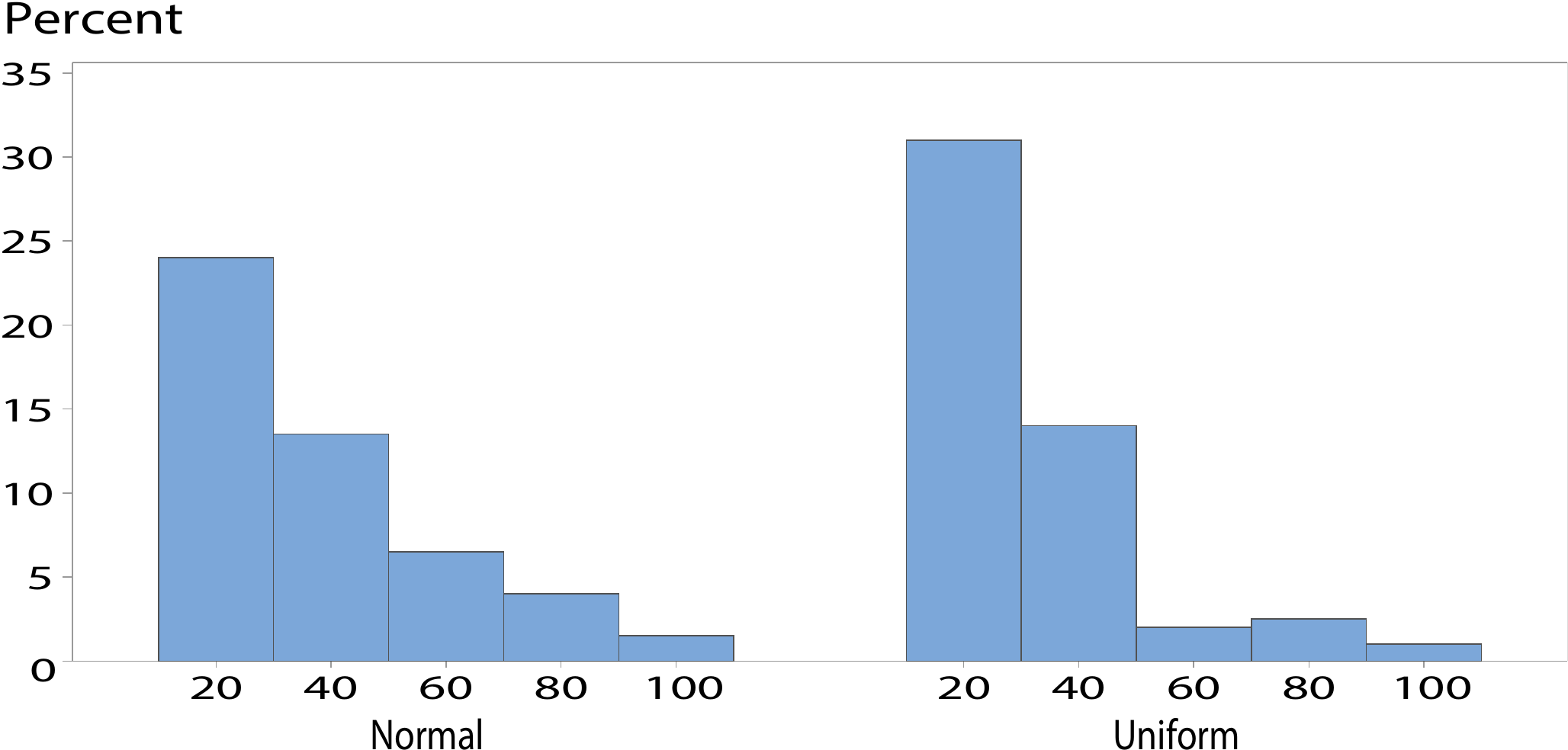}\\
 \caption{Percent of times {\CP}$_{10}$ equals {\CP}$_{30}$ by size.}\label{fig:barchart}
\end{figure}

\begin{figure*}
  \centering
  \includegraphics[width=\textwidth,keepaspectratio]{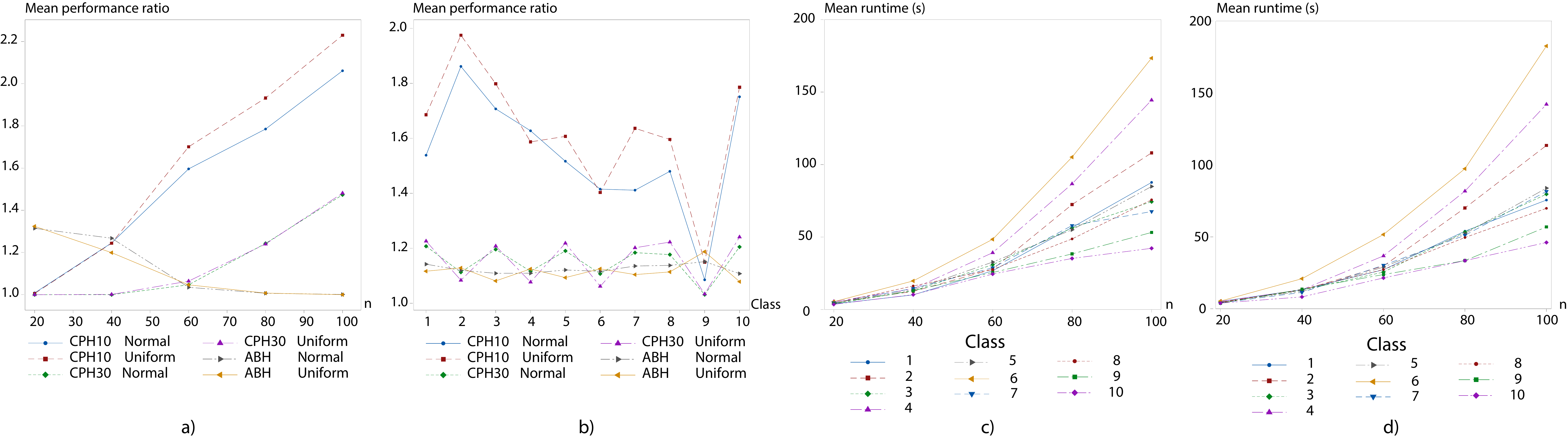}\\
  \caption{Mean performance ratio as a function of size (a) and class (b). Mean runtime (s) of {\AB} for instances with normally (c) and uniformly distributed (d) due dates.}\label{fig:Figure2}
\end{figure*}

Figure \ref{fig:results} displays the heat map of the performance ratio $\frac{z_H}{z_{best}},\ H \!\in\! \mathcal{H}\!=\!\{${\CP}$_{10},$ {\CP}$_{30},$ {\AB}$\},$ where {\CP}$_{10}$ and {\CP}$_{30}$ denote {\CP} when allocated 10 and 30 minutes of runtime respectively, $z_{best}=\min_{H \in \mathcal{H}}\{z_H\},$ and a dark blue color signals a better objective function value.  It suggests that allocating a longer runtime to {\CP} enhances its mean performance ratio; in particular when $n>20.$ This enhancement occurs in about 80\% of instances. The percent of times {\CP}$_{10}$ equals {\CP}$_{30}$ decreases as $n$ increases, as illustrated by Figure \ref{fig:barchart}. This is further supported by a paired statistical test and by Figure \ref{fig:Figure2}.a-b, which displays the mean performance ratios as a function of size and class. It is unlikely that increasing the runtime of CPH beyond 30 minutes for larger instances will induce further improvements shortly because of the large size of the search space.
{\AB} outperforms {\CP} for instances with more than 60 items whereas {\CP} performs better than {\AB} when the number of items is 20 or 40.  For $n=60$, {\AB} outperforms {\CP} for set $\mathcal{L}$ while being outperformed by {\CP} for set $\mathcal{S}.$  {\CP} investigates a substantially higher portion of the solution space when the search space is small than when the search space is large. As $n$ increases, {\CP} struggles to reduce the search space (thus, to converge to low-penalty solutions) whereas {\AB} reaches such solutions for the larger instances progressively thanks to its general rules of thumb that aim at minimizing the incremental weighted earliness tardiness as {\AB} assigns items to bins.

Analysis of variance tests further show that the mean performance ratio is not sensitive to the distribution of the due dates but is sensitive to both the class and the problem size.  For single machine scheduling, problems with normally distributed due dates are harder to solve; however, for the problem at hand, clusters of items with very close due dates can be scheduled in the same bin. On the other hand, processing items with uniformly distributed due dates in a single batch increases the objective function value, and constitutes an additional difficulty to the problem. This is reflected by \textbf{slightly} lower solution quality of {\CP} and {\AB} and \textbf{slightly} larger run times.

A hypothesis test infers that there is not enough statistical evidence to claim that the mean runtime of {\AB} is larger when the due dates are normally distributed than when they are uniformly generated. A one-way analysis of variance further confirms that the runtime of {\AB} is not sensitive to the distribution of due dates.  That is, processing the items in batches desensitizes {\AB} to the clustering of the due dates. The mean runtime of {\AB} is on the other hand dependent on the problem size and class as illustrated by Figure \ref{fig:Figure2}.c-d, which shows the main interaction of size, class, and distribution of due dates on the mean runtime of {\AB}. The maximal mean runtime of 203 s of {\AB} is far less than the allocated 10 minutes of {\CP}$_{10}$.
The larger the bin to item size, the larger the run time of {\AB} is. This is particularly apparent for larger instances of classes 2, 4, 6 and 10, which all have high bin to item size ratios. As the number of items per bin increases, the number of potential arrangements of items increases too; making it harder to determine whether the packing of a given item set to the bin is feasible.  Unfortunately, increasing the threshold of the runtime of the constraint program {\pack} did not enhance the performance of {\AB} for those classes of instances. Indeed, the feasibility problem remains too hard to solve when solutions lay ``on the edge of feasibility''.
\begin{figure}
  \centering
  \includegraphics[width=\columnwidth,height=4cm]{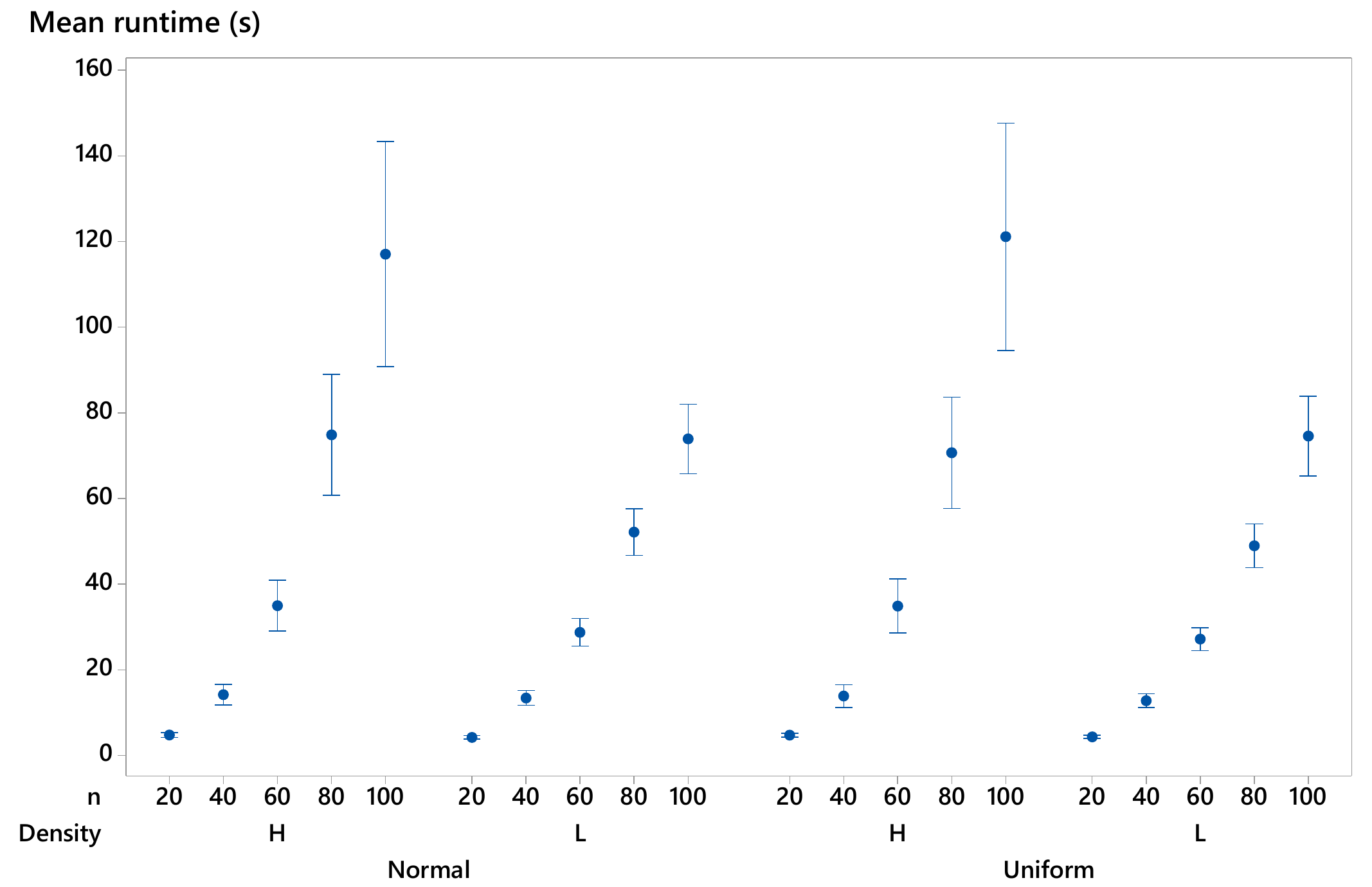}\\
  \caption{95\% confidence intervals for the mean runtime (s) of {\AB} for instances with low and high density bins.}\label{fig:RTABsets}
\end{figure}
Figure \ref{fig:RTABsets}, which displays the 95\% confidence intervals of the mean runtime (s) of {\AB} as a function of size for sets $\mathcal{L}$ and $\mathcal{S},$ provides further computational proof that {\AB} runs faster on the instances with relatively large items. 

{\AB} performs consistently better for instances with low bin to item size ratios than for instances with high ratios. This is most likely due to the ``ripple effect'', which arises from low-penalty items forcing items with higher penalties away from their most proximal bins. While the search for ideal processing windows favors the assignment of items to their most proximal bins, the asymmetric nature of the per unit penalties $\epsilon$ and $\tau$  does not guarantee low earliness tardiness to all items assigned to a bin. As filled bins near their full capacity, the greedy {\GroupJoin} and {\GroupFormation} actions may result in locally-optimal scenarios where small, low-penalty items prohibit large, high-penalty items from being packed into their proximal bins; thus, force them into bins that are far away from their due dates. This occurs more often in instances with high bin to item ratios because bins can pack more items (which are not necessarily homogeneous).

\section{Conclusions}\label{sec:Conclusion}
This paper introduces a very pertinent industrial problem that combines two very hard combinatorial optimization problems: a two-dimensional guillotine bin packing / cutting, and a single machine weighted earliness tardiness batch scheduling.  It defines the problem, and models it using two different approaches: one based on constraint programming and one based on agent based modeling where items and bins act as cooperating negotiating agents. The computational results on randomly generated instances that will serve as benchmark sets for future research show that the constraint programming approach is more promising when the problem size is small while the agent based model gives better results for larger instances thanks to the common sense rules that govern the negotiations of the agents. 

The first heuristic can be enhanced with additional selection / dominance rules that will prune large parts of the search space while the second can be improved with a neighborhood search.  In addition to improving these two heuristics, future research can focus on applying and adapting enumerative approaches, such as genetic algorithms, to the specificities of this problem. In fact, it would be interesting to design/adapt a high-performing evolutionary technique that can cope with the large number of infeasible solutions it would encounter during its search.  Moreover, the problem can be extended to more complex manufacturing set ups such as flow shops and job shops, to cutting problems with more complex shapes and constraints as in apparel manufacturing, to three dimensional shapes as in transportation, or to the variable-sized variable-cost bin packing problem.

\section*{Acknowledgments}

This work has been supported by the ARC Discovery Project DP130104395.

\bibliographystyle{model5-names}
\bibliography{references}

\begin{thebibliography}{20}
\expandafter\ifx\csname natexlab\endcsname\relax\def\natexlab#1{#1}\fi
\providecommand{\url}[1]{\texttt{#1}}
\providecommand{\href}[2]{#2}
\providecommand{\path}[1]{#1}
\providecommand{\DOIprefix}{doi:}
\providecommand{\ArXivprefix}{arXiv:}
\providecommand{\URLprefix}{URL: }
\providecommand{\Pubmedprefix}{pmid:}
\providecommand{\doi}[1]{\href{http://dx.doi.org/#1}{\path{#1}}}
\providecommand{\Pubmed}[1]{\href{pmid:#1}{\path{#1}}}
\providecommand{\bibinfo}[2]{#2}
\ifx\xfnm\relax \def\xfnm[#1]{\unskip,\space#1}\fi
\bibitem[{Arbib \& Marinelli(2014)}]{Arbib14}
\bibinfo{author}{Arbib, C.}, \& \bibinfo{author}{Marinelli, F.}
  (\bibinfo{year}{2014}).
\newblock \bibinfo{title}{On cutting stock with due dates}.
\newblock {\it \bibinfo{journal}{Omega}\/},  {\it \bibinfo{volume}{46}\/},
  \bibinfo{pages}{11 -- 20}.
\bibitem[{BenBassat(2016)}]{plataine}
\bibinfo{author}{BenBassat, M.} (\bibinfo{year}{2016}).
\newblock \bibinfo{title}{Production chain optimization for wood frame
  manufacturers}.
\newblock {\it
  \bibinfo{journal}{{www.pla\-tai\-ne.com/\-pro\-duc\-tion-chain-\-op\-ti\-mi\-za\-tion-\-for-\-wood-\-fra\-me-ma\-nu\-fac\-tu\-rers}}\/},
  .
\newblock \bibinfo{note}{[Online; accessed December 2016]}.
\bibitem[{Bockmayr \& Hooker(2005)}]{Bockmayr05}
\bibinfo{author}{Bockmayr, A.}, \& \bibinfo{author}{Hooker, J.~N.}
  (\bibinfo{year}{2005}).
\newblock \bibinfo{title}{Constraint programming}.
\newblock In \bibinfo{editor}{G.~N. K.~Aardal}, \&
  \bibinfo{editor}{R.~Weismantel} (Eds.), {\it \bibinfo{booktitle}{Discrete
  Optimization}\/} (pp. \bibinfo{pages}{559 -- 600}).
\newblock \bibinfo{publisher}{Elsevier} volume~\bibinfo{volume}{12} of {\it
  \bibinfo{series}{Handbooks in Operations Research and Management Science}\/}.
\bibitem[{Bonyadi \& Michalewicz(2016)}]{Bonyadi2016}
\bibinfo{author}{Bonyadi, M.~R.}, \& \bibinfo{author}{Michalewicz, Z.}
  (\bibinfo{year}{2016}).
\newblock \bibinfo{title}{Evolutionary computation for real-world problems}.
\newblock In {\it \bibinfo{booktitle}{Challenges in Computational Statistics
  and Data Mining}\/} (pp. \bibinfo{pages}{1--24}).
\bibitem[{Burke et~al.(2012)Burke, Hyde, Kendall \& Woodward}]{Burke2012}
\bibinfo{author}{Burke, E.~K.}, \bibinfo{author}{Hyde, M.~R.},
  \bibinfo{author}{Kendall, G.}, \& \bibinfo{author}{Woodward, J.}
  (\bibinfo{year}{2012}).
\newblock \bibinfo{title}{Automating the packing heuristic design process with
  genetic programming}.
\newblock {\it \bibinfo{journal}{Evol. Comput.}\/},  {\it
  \bibinfo{volume}{20}\/}, \bibinfo{pages}{63--89}.
\bibitem[{Fekete \& Schepers(2004)}]{Fekete04}
\bibinfo{author}{Fekete, S.~P.}, \& \bibinfo{author}{Schepers, J.}
  (\bibinfo{year}{2004}).
\newblock \bibinfo{title}{A general framework for bounds for higher-dimensional
  orthogonal packing problems.}
\newblock {\it \bibinfo{journal}{Math. Meth. of OR}\/},  {\it
  \bibinfo{volume}{60}\/}, \bibinfo{pages}{311--329}.
\bibitem[{Haz{\i}r \& Kedad-Sidhoum(2014)}]{Hazır2012}
\bibinfo{author}{Haz{\i}r, {\"O}.}, \& \bibinfo{author}{Kedad-Sidhoum, S.}
  (\bibinfo{year}{2014}).
\newblock \bibinfo{title}{Batch sizing and just-in-time scheduling with common
  due date}.
\newblock {\it \bibinfo{journal}{Annals of Operations Research}\/},  {\it
  \bibinfo{volume}{213}\/}, \bibinfo{pages}{187--202}.
\bibitem[{Hooker(2002)}]{H02}
\bibinfo{author}{Hooker, J.~N.} (\bibinfo{year}{2002}).
\newblock \bibinfo{title}{Logic, optimization, and constraint programming}.
\newblock {\it \bibinfo{journal}{INFORMS Journal on Computing}\/},  {\it
  \bibinfo{volume}{14}\/}, \bibinfo{pages}{295 -- 321}.
\bibitem[{Lodi et~al.(2014)Lodi, Martello, Monaci \& Vigo}]{Lodi2014107}
\bibinfo{author}{Lodi, A.}, \bibinfo{author}{Martello, S.},
  \bibinfo{author}{Monaci, M.}, \& \bibinfo{author}{Vigo, D.}
  (\bibinfo{year}{2014}).
\newblock \bibinfo{title}{Two-dimensional bin packing problems}.
\newblock In {\it \bibinfo{booktitle}{Paradigms of Combinatorial
  Optimization}\/} (pp. \bibinfo{pages}{107--129}).
\newblock \bibinfo{publisher}{John Wiley \& Sons, Inc.}
\bibitem[{Lodi et~al.(2002)Lodi, Martello \& Vigo}]{Lodi2002379}
\bibinfo{author}{Lodi, A.}, \bibinfo{author}{Martello, S.}, \&
  \bibinfo{author}{Vigo, D.} (\bibinfo{year}{2002}).
\newblock \bibinfo{title}{Recent advances on two-dimensional bin packing
  problems}.
\newblock {\it \bibinfo{journal}{Discrete Applied Mathematics}\/},  {\it
  \bibinfo{volume}{123}\/}, \bibinfo{pages}{379 -- 396}.
\bibitem[{M'Hallah \& Alhajraf(2015)}]{Hallah2015}
\bibinfo{author}{M'Hallah, R.}, \& \bibinfo{author}{Alhajraf, A.}
  (\bibinfo{year}{2015}).
\newblock \bibinfo{title}{Ant colony systems for the single-machine total
  weighted earliness tardiness scheduling problem}.
\newblock {\it \bibinfo{journal}{Journal of Scheduling}\/},  (pp.
  \bibinfo{pages}{1--15}).
\bibitem[{Polyakovskiy et~al.(2014)Polyakovskiy, Bonyadi, Wagner, Michalewicz
  \& Neumann}]{Polyakovskiy2014TTP}
\bibinfo{author}{Polyakovskiy, S.}, \bibinfo{author}{Bonyadi, M.~R.},
  \bibinfo{author}{Wagner, M.}, \bibinfo{author}{Michalewicz, Z.}, \&
  \bibinfo{author}{Neumann, F.} (\bibinfo{year}{2014}).
\newblock \bibinfo{title}{A comprehensive benchmark set and heuristics for the
  traveling thief problem}.
\newblock In {\it \bibinfo{booktitle}{Proceedings of the 2014 Annual Conference
  on Genetic and Evolutionary Computation}\/} GECCO '14 (pp.
  \bibinfo{pages}{477--484}).
\newblock \bibinfo{address}{New York, NY, USA}: \bibinfo{publisher}{ACM}.
\bibitem[{Polyakovskiy \& M'Hallah(2011)}]{Polyakovskiy11}
\bibinfo{author}{Polyakovskiy, S.}, \& \bibinfo{author}{M'Hallah, R.}
  (\bibinfo{year}{2011}).
\newblock \bibinfo{title}{An intelligent framework to online bin packing in a
  just-in-time environment}.
\newblock In {\it \bibinfo{booktitle}{Modern Approaches in Applied
  Intelligence}\/} (pp. \bibinfo{pages}{226--236}).
\newblock \bibinfo{publisher}{Springer} volume \bibinfo{volume}{6704} of {\it
  \bibinfo{series}{LNCS}\/}.
\bibitem[{Polyakovskiy \& M'Hallah(2014)}]{Polyakovskiy2014115}
\bibinfo{author}{Polyakovskiy, S.}, \& \bibinfo{author}{M'Hallah, R.}
  (\bibinfo{year}{2014}).
\newblock \bibinfo{title}{A multi-agent system for the weighted earliness
  tardiness parallel machine problem}.
\newblock {\it \bibinfo{journal}{Computers \& Operations Research}\/},  {\it
  \bibinfo{volume}{44}\/}, \bibinfo{pages}{115 -- 136}.
\bibitem[{Polyakovsky \& M'Hallah(2009)}]{Polyakovsky2009767}
\bibinfo{author}{Polyakovsky, S.}, \& \bibinfo{author}{M'Hallah, R.}
  (\bibinfo{year}{2009}).
\newblock \bibinfo{title}{An agent-based approach to the two-dimensional
  guillotine bin packing problem}.
\newblock {\it \bibinfo{journal}{European Journal of Operational Research}\/},
  {\it \bibinfo{volume}{192}\/}, \bibinfo{pages}{767 -- 781}.
\bibitem[{Refalo(2004)}]{Refalo2004}
\bibinfo{author}{Refalo, P.} (\bibinfo{year}{2004}).
\newblock \bibinfo{title}{Impact-based search strategies for constraint
  programming}.
\newblock In {\it \bibinfo{booktitle}{Principles and Practice of Constraint
  Programming -- CP 2004}\/} (pp. \bibinfo{pages}{557--571}).
\newblock \bibinfo{publisher}{Springer}.
\bibitem[{Reinertsen \& Vossen(2010)}]{Reinertsen10}
\bibinfo{author}{Reinertsen, H.}, \& \bibinfo{author}{Vossen, T.~W.}
  (\bibinfo{year}{2010}).
\newblock \bibinfo{title}{The one-dimensional cutting stock problem with due
  dates}.
\newblock {\it \bibinfo{journal}{European Journal of Operational Research}\/},
  {\it \bibinfo{volume}{201}\/}, \bibinfo{pages}{701 -- 711}.
\bibitem[{Sim et~al.(2015)Sim, Hart \& Paechter}]{Sim201537}
\bibinfo{author}{Sim, K.}, \bibinfo{author}{Hart, E.}, \&
  \bibinfo{author}{Paechter, B.} (\bibinfo{year}{2015}).
\newblock \bibinfo{title}{A lifelong learning hyper-heuristic method for bin
  packing}.
\newblock {\it \bibinfo{journal}{Evolutionary Computation}\/},  {\it
  \bibinfo{volume}{23}\/}, \bibinfo{pages}{37--67}.
\bibitem[{Verstichel et~al.(2015)Verstichel, Kinable, Causmaecker \&
  Berghe}]{Verstichel15}
\bibinfo{author}{Verstichel, J.}, \bibinfo{author}{Kinable, J.},
  \bibinfo{author}{Causmaecker, P.~D.}, \& \bibinfo{author}{Berghe, G.~V.}
  (\bibinfo{year}{2015}).
\newblock \bibinfo{title}{{A Combinatorial Benders' decomposition for the lock
  scheduling problem}}.
\newblock {\it \bibinfo{journal}{Computers \& Operations Research}\/},  {\it
  \bibinfo{volume}{54}\/}, \bibinfo{pages}{117 -- 128}.
\bibitem[{Yager \& Filev(1984)}]{Yager}
\bibinfo{author}{Yager, R.}, \& \bibinfo{author}{Filev, D.}
  (\bibinfo{year}{1984}).
\newblock {\it \bibinfo{title}{Essentials of Fuzzy Modeling and Control}\/}.
\newblock \bibinfo{publisher}{John Wiley}.

\end{thebibliography}

\end{document}